\documentclass{aircc}
\usepackage{mathpartir}
\usepackage{cite}
\usepackage{hyperref}
\usepackage{amsmath,amssymb,amsfonts}
\usepackage{algorithmic}
\usepackage{graphicx}
\usepackage{textcomp}
\usepackage{xcolor}
\usepackage{tikz}
\usepackage{listings}
\usepackage{float}
\usepackage{wrapfig}

\newcommand*\circled[1]{\tikz[baseline=(char.base)]{
            \node[shape=circle,draw,inner sep=0.7pt] (char) {#1};}}
\begin{document}

\title{
From Monolith to Microservices: Software Architecture for Autonomous UAV Infrastructure Inspection
}

\author{Lea Matlekovic and Peter Schneider-Kamp}
\affiliation{Department of Mathematics and Computer Science, University of Southern Denmark, Odense, Denmark \\ \email{\url{matlekovic@imada.sdu.dk}} \\ \email{\url{petersk@imada.sdu.dk}}}

\maketitle
\begin{abstract}
Linear-infrastructure Mission Control (LiMiC) is an application for autonomous Unmanned Aerial Vehicle (UAV) infrastructure inspection mission planning developed in monolithic software architecture. The application calculates routes along the infrastructure based on the users' inputs, the number of UAVs participating in the mission, and UAVs' locations. LiMiC1.0 is the latest application version migrated from monolith to microservices, continuously integrated, and deployed using DevOps tools to facilitate future features development, enable better traffic management, and improve the route calculation processing time. Processing time was improved by refactoring the route calculation algorithm into services, scaling them in the Kubernetes cluster, and enabling asynchronous communication in between. In this paper, we discuss the differences between the monolith and microservice architecture to justify our decision for migration. We describe the methodology for the application's migration and implementation processes, technologies we use for continuous integration and deployment, and we present microservices improved performance results compared with the monolithic application.
\end{abstract}


\section{Introduction}

Infrastructure inspection is a dangerous, expensive, and time-consuming task. Falls from inspection sites are the leading cause of death among construction workers \cite{b1}. It is not the only risk construction workers face. They may also come into contact with toxic chemicals, moving machinery, speeding traffic, or high-voltage equipment. The risks increase when the infrastructure is in disrepair. Deploying UAVs to inspection eliminates the safety risks. Defects can be detected from the UAV camera and construction workers can see the state of an inspection site before they climb and start repairing the infrastructure. The UAV technology reduces the costs of inspection since it does not require special inspection equipment, helicopters nor airplanes \cite{b2}. Nowadays, many companies use UAVs for visual infrastructure inspection, but they still control them manually or use automated flying. By increasing the degree of UAVs' autonomy, we expect to reduce the costs even more and save time on process optimization.
To develop an autonomous system for infrastructure inspection, we need to plan the inspection mission and schedule tasks for each UAV participating in the mission. 

The preliminary system design for autonomous UAV infrastructure inspection was described in \cite{b3}. The system was designed in three layers: cloud services, UAVs, and communication between them. Global mission planning and scheduling software is deployed in the cloud. When the mission is calculated, the route coordinates are sent to the UAVs. Message exchange between UAVs and cloud services is through the HTTP (Hypertext Transfer Protocol). Robot Operating System (ROS) runs on UAVs where high-level control software is deployed. PX4 open-source software is used as a low-level flight controller.

Linear-infrastructure Mission Control (LiMiC) is presented in \cite{b4}. It is a software application for global mission planning and scheduling developed in a monolithic architecture. Monolithic means that all the application logic was composed into a single program. To facilitate further development and improve the route calculation processing time, we decided to redesign the architecture and compare the performance.

The paper is structured as follows. In \hyperref[sec:two]{Section 2}, we compare monolithic architecture to microservices, justifying our choice for a redesign. We describe DevOps practices that facilitate microservices deployment to the production. Related work, concerning monolith decoupling is presented in \hyperref[sec:three]{Section 3}. In \hyperref[sec:four]{Section 4}, we describe LiMiC application logic, architecture redesign based on LiMiC functionalities, migration and implementation processes, technologies we used as well as the deployment strategy. In \hyperref[sec:five]{Section 5}, we compare performance between the application developed in monolith and microservice architecture. We conclude in \hyperref[sec:six]{Section 6} discussing possible improvements and future work.     

\section{Preliminaries}
\label{sec:two}
In this section, we describe software architecture styles relevant for this article and present their benefits as well as challenges. Software architecture in general describes application organization and structure. Architectural decisions impact application quality, performance, maintainability, and usability. Software architecture chosen at the beginning of the application development can have a high impact on the application's future and affect its success. Architecture should be considered and planned if the application needs to be scalable, maintainable, and easily upgradable. When developing the application from scratch, the focus is usually on having a working product as soon as possible. However, that approach can become unsustainable when the number of application features grows fast. If the architecture is not reconsidered, the development slows down and application becomes difficult to maintain. 

\subsection{Monolithic Architecture}
Monolithic applications encompass several tightly coupled functions and tend to have a big codebase. The development does not require advanced architecture planning since the application is developed, packaged, and deployed as a standalone instance. The packaged application can be deployed to the server and scaled horizontally by running multiple instances behind a load balancer. The load balancer distributes the traffic across the instances deployed on the different servers. Testing the monolith applications can be performed end-to-end by launching the application and using existing testing framework e.g., Selenium \cite{b22}... However, when the code base grows and many developers work on the same application, monolith applications face challenges. The application can become too complex and difficult to understand. That prevents quick updates and development slows down. On each update, the entire application must redeploy and it can be difficult to track update impacts. It leads to extensive and time-consuming manual testing. A bug in any module can potentially bring the whole system down since all instances rely on the same code base. Even though the application can be scaled, the load is usually not equally distributed to all the modules i.e., the traffic is directed to only one application module, but they are all scaled equally. When application modules have different resource requirements, like they often do, it can lead to unnecessary CPU (Central Processing Unit) and memory consumption. Monoliths are not robust to changes and adopting a new framework or technology would lead to rewriting of the entire code base which is both expensive and time-consuming \cite{b5}. Development in monolithic architecture could be a good choice if the application will not need to be further extended or as a starting point when the main goal is to have a simple, working end-product. However, when the development slows down as the codebase grows, it is recommended to reconsider the architecture \cite{b26}.          

\subsection{Microservice Architecture}
Microservice architecture, also known as Microservices, is a software architecture that structures an application as a collection of small, loosely coupled services. The services are independently deployable, highly maintainable, and testable \cite{b7}. The concept was developed to overcome the downsides of monolithic architecture. Microservices have clear boundaries between each other and communicate through the HTTP protocol, usually by exposing a REST API and sending requests. Each service represents one capability that makes it easier to understand and locate the code. It also makes them robust to changes. Since they are small and deployed independently, they are easy to update and maintain. Services can be scaled independently and automatically, depending on the load. They can use different technology stacks, including programming language and data storage. That gives high flexibility to the development teams. However, there are some drawbacks of microservice architecture. The fact that a microservices application is a distributed system requires handling of fallacies the distributed computing carries. It means that developers must deal with the additional complexity.  

There are opinions suggesting to start application development in monolith architecture first and then migrate to microservices \cite{b21}. As monolithic systems become too large to deal with, many enterprises are drawn to breaking them down into microservices. On the other side, others recommend starting with microservices if that architectural style is the goal \cite{b6}. However, most large scale websites including Netflix, Amazon, and eBay have evolved from a monolithic architecture to microservices \cite{b8}. 

\subsection{DevOps}
 Microservices, since they are independent, bring more complexity in application deployment than monolithic applications. There are many practices and tools developed to facilitate testing, integration, and deployment of microservices \cite{b9}. DevOps is a combination of practices and tools designed to facilitate the delivery of applications. It aims to increase an organization's ability to deploy applications faster by removing the barriers between development and operations teams \cite{b10}. DevOps practices automatize the delivery process and are implemented as a part of a production pipeline. Applied tools and practices depend on the application delivery requirements and goals. Continuous integration is the practice of merging changes to the main branch as often as possible. When developers commit local changes to the remote repository, automated build and tests can run there before proceeding to production. Remote repository platforms with built-in version control, like GitLab and GitHub, facilitate the collaboration between developers and enable continuous integration. These platforms also integrate with different DevOps tools to enable continuous delivery and deployment. The practice of continuous delivery includes continuous integration and after a successful build, automatically propagates the application to staging. In staging, the application is deployed to the testing environment. There, the application including all services can be run and tested. Continuous delivery requires manual approval for release into production. However, the continuous deployment includes all the steps described in continuous integration and delivery, but instead of manual, the release process is also automated. Described process can vary in complexity depending on the number of services, test requirements, and in general, the deployment strategy.

\section{Related work}
\label{sec:three}
The concept of microservices arose around ten years ago and today is a widely used concept for large enterprises to develop their software systems. According to IDC (International Data Corporation), by 2022, 90\% of all new applications will be based on microservices architectures \cite{b26}. Adopting microservices improves agility and flexibility, enabling enterprises to bring their products and services to market faster. 

Although the benefits of microservices are evident, adopting it is not an easy task as it usually involves refactoring the monolithic application. There are many critical questions to ask before deciding to refactor the monolith. Will the refactoring bring value? How to re-architect an existing system without having to stop all other work on it? How big should a microservice be? What are some of the migration patterns you could adopt when splitting up a monolith? \cite{b26} There is some research tackling these questions in the latest years. Most of the previous research on microservices either identifies challenges when splitting the monolith like in \cite{b27}, or proposes refactoring methods. Classification of refactoring approaches is presented in \cite{b28}. There are four notable approaches identified. Static Code Analysis approaches require the application’s source code analysis and derive a decomposition from it through possible intermediate stages. Meta-Data aided approaches require more abstract input data, like architectural descriptions in form of UML (Unified Modeling Language) diagrams, use cases, interfaces, or historical VCS (Version Control System) data. Workload-Data aided approaches aim to find suitable service cuts by measuring the application’s operational data (e.g. communication, performance) on module or function level and use this data to determine a fitting decomposition and granularity. Dynamic Microservice Composition approaches try to solve the problem more holistically by describing a microservices runtime environment. Other than the previously mentioned categories, the resulting set of services is permanently changing in each iteration of re-calculating the best-fitting composition (based on e.g. workload). Static and dynamic analysis of the monolith is further described and used for refactoring in \cite{b29}. The authors used dynamic software visualization to identify appropriate microservice boundaries for a real-world application. A Dataflow-driven approach is proposed in \cite{b30}. The article presents a semi-automatic mechanism to break business logic into microservices and visualize services with data flows. Although the refactoring algorithm could save some time in splitting the monolith, it still requires identification and description of monolith application business logic. All the research work we encountered based refactoring on application logic and we followed the same approach. Another interesting work is proposed in \cite{b31} where the focus is on profitability depending on the number of refactored services and deployment. The article shares lessons learned on an industrial migration to a web-oriented microservice architecture. The services are refactored based on capabilities and the decision on service size is based on the company organization and profitability. For deployment tools, Docker is emphasized as a widely used container technology for achieving service isolation. The article presents the return of investment from refactoring the monolith and concludes by presenting the benefits of the transition for a given company.

Even though the previous research literature guided us in choosing the most suitable approaches, each application is different and requires an individual approach bringing new challenges when splitting it into microservices.

\section{Methodology}
\label{sec:four}
In this section, we describe processes leading to LiMiC migration from monolith to microservice architecture. Migration from monolith to microservices is not an easy task \cite{b16}. There are many different approaches to splitting the monolith. The code should be studied before decoupling into the logical components. Dependencies should be identified and isolated. Even though our monolithic application was not deployed in the production and it did not have users depending on it, we encountered difficulties while refactoring, and parts of the code had to be rewritten instead of reused.

Main processes leading to LiMiC1.0 implementation from monolith to microservices are as follows:
\begin{itemize}
    \item \textbf{Identification} - identification of features in monolithic application
    \item \textbf{Isolation} - isolation of features where each isolated feature is one meaningful and standalone unit 
    \item \textbf{Implementation} - implementation of isolated features withing framework enabling the communication between them
\end{itemize}

For application deployment we developed a strategy and used DevOps tools described in this section to implement and automatize processes leading to deployment. By enabling continuous integration and continuous deployment, we tremendously benefited from the microservice architecture, i.e. better codebase organization, faster development, easier integration and deployment.

\subsection{Identification \& Isolation}
Linear-infrastructure Mission Control (LiMiC) is an application software for autonomous Unmanned Aerial Vehicle (UAV) infrastructure inspection. It is a mission planner for calculating the near-optimal routes for power line inspection using UAVs. The user can choose the mission targets, i.e. power towers for inspection, calculate the order of visiting and generate waypoints. Waypoints represent the coordinates sent to the UAVs for navigation to the inspection targets.

We used data from the OpenStreetMap \cite{b11} to represent the power towers on the web interface and to build a graph using NetworkX Python package \cite{b12}. The web interface shows UAVs and towers on a 2D map. We used graph structure to describe and store path data. Nodes in the graph represent power towers. Nodes contain the tower's unique identifier and geographical location. Edges between the nodes contain distances between the neighboring towers which we used to describe the costs of flying from one tower to another. We used Google OR-Tools \cite{b13} to find the near-optimal visiting schedule for multiple vehicles visiting a set of inspection targets stored in the graph. When a user selects inspection targets, the vehicle scheduling algorithm generates a distance matrix using the A* algorithm. A* finds the shortest path, based on the distance data stored in the graph, for each combination of UAVs and towers participating in the mission and returns it to the vehicles scheduler. Calculated paths are used for determining the visiting schedule, i.e. which UAV will inspect which tower. The result is visualized on the map and paths are stored to be sent to the UAVs. Paths are represented as a set of waypoints containing geographical locations. 

\begin{figure}[htbp]
\centerline{\includegraphics[scale=0.45]{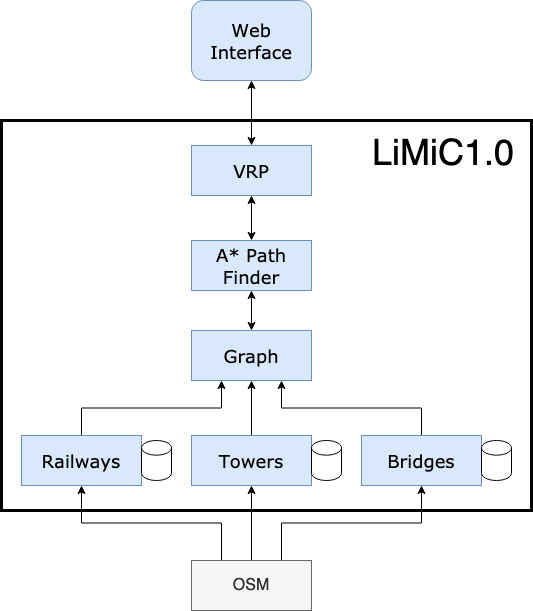}}
\caption{Application structure based on capabilities}
\label{fig}
\end{figure}

We developed the application in monolithic architecture structured as Python modules situated in a folder. The frontend communicates with the backend through HTTP requests using Flask Python framework \cite{b14}. As the application was growing and many developers started contributing, we started thinking about redesigning the structure. We downloaded the data from the OpenStreetMap and saved the graph locally. If we wanted to update the graph we would need to run the graph generation manually and save a new one. Since we plan to grow the application and deploy it to production, it was necessary to redesign the graph creation and update it automatically in defined time intervals. The codebase was not organized in a logical structure and it was difficult to navigate. Addition of a new feature required searching through the codebase to understand the processes before contributing to the code. A lot of time that could be spent on the development was invested in understanding the existing implementation. We wanted to facilitate the feature addition and reduce the time spent on understanding the code. That required a fixed structure, but flexible and adaptive to feature addition. In order to grow the application and deploy it to production, we decided to restructure it. Another concern was the load distribution. Some parts of the application are more static while others are heavily used. If deployed to the production, some parts of the application would have to be unnecessarily scaled. Therefore, we structured the application in microservices based on the application's capabilities. The proposed structure is shown in Figure \ref{fig}. The arrows determine the communication flow between the services, e.g. the graph service sends a request to the towers service while the towers service provides power tower information needed to create a graph. The tower service should contain the capability to extract the data from the OpenStreetMap, organize it in a suitable form and save it to the database. The same capabilities should exist for other infrastructures like bridges and railways. Bridges can be reached by calculating the route near the power lines or railways leading to the bridge location. Depending on the infrastructure we want to inspect, we would build the graph containing nodes with locations of power towers, railway towers, or bridges as defined in the OpenStreetMap. The data is used by the web interface to visualize the inspection targets. Graph service would then use available data to create a graph. Inspection targets chosen on the web interface would be sent to the vehicle routing problem (VRP) solver service, which uses the A* algorithm to find the shortest path between the targets and UAVs and determines the routing schedule. A* calculates the shortest path based on the graph created in the graph service.

\subsection{Implementation}
 In the previous subsection, we identified services logically, based on the capabilities of the monolithic application. In this subsection, we describe the process of monolith decoupling and implementation in detail. 
 
 The decoupling process was divided into two parts. First was the graph creation based on the data stored in a database created in railways, towers, and bridges services. That way, we were able to test the generated graph file with a monolithic application. The second part was splitting the routing solver and A* pathfinder into independent components. Every time the request comes to the routing solver, it requests the shortest path for each set of targets and UAV locations from the A*. A* requests a graph from the graph service only on the first request. All the services use FastAPI \cite{b17} web framework for exposing APIs (Application Programming Interfaces) to the web interface or other services. FastAPI enables input validation where we define how the input is structured. It facilitates the input usage in the code without installing additional packages. FastAPI provides interactive, automatically generated documentation which we find very useful for endpoint visualization and testing. When looking into the performance comparison, FastAPI is much faster than Flask which was used for LiMiC monolith development \cite{b20}. It also provides asynchronous support crucial for taking advantage of microservice architecture. Asynchronous requests make the communication between routing solver and A* pathfinder much faster, especially when A* pathfinder is scaled. It enables concurrent computation of the shortest path.     

Services are developed as follows:
\begin{itemize}
\item \textit{Towers service} - the service uses Overpass API \cite{b15} to get the power towers and power lines data from OpenStreetMap. Service connects to the MongoDB Atlas database service and creates a database if it already does not exist. MongoDB Atlas provides a cloud database service for MongoDB databases. We chose MongoDB because it stores data in JSON-like documents and can be easily queried. Furthermore, it provides a geo-based search we use for finding indirect towers' neighbors needed for creating edges in the graph. Power towers are stored as a collection of objects with a unique identifier and location containing the tower's coordinates. Power lines are extracted using the same method and stored as a collection of objects with unique identifiers and an array of unique identifiers representing the nodes the line is passing through. Most of the nodes' unique identifiers are also towers', but they can also be a line intersections. For that reason, we created another collection of nodes contained in the power line arrays. Nodes' objects are defined the same as towers. Power line collection also contains tags with useful information on the number of cables, frequency, and voltage. The need for storing the power line information emerged because we want to find towers' direct neighbors. Tower's direct neighbor is the closest tower, or more of them, laying on the same power line. When creating the graph we want to give a higher cost to the paths between the indirect neighbors than direct since we want the UAVs to fly close to the infrastructure whenever that is possible. Such a strategy is implemented for UAV flight regulations reasons. Tower service exposes two endpoints. One is providing the tower data and another one the data containing power lines. The data is used by the web interface to visualize the inspection targets. Graph service requests the data to create nodes and edges in the graph. 
\item \textit{Railways service} - the service extracts railways location in the same way described in the towers service. It saves the data to the MongoDB database and sends it to the graph service on request.  
\item \textit{Bridges service} - the service extracts polygons around the bridges and saves them to the database. Polygons are stored as the way types containing unique identifiers for nodes in the polygon. The nodes are stored in the same database containing the node's location. Bridges locations than can be combined with power line locations to create a graph and enable the UAVs to reach a bridge.   
\item \textit{Graph service} - the service creates a graph based on the data received from either towers service or railways and bridges. The towers' locations represent the nodes in the graph while a pair of neighboring towers represent edges. The edge's cost is set up as the calculated distance between the neighboring towers. All the data for building the graph is requested from the towers service. The graph is created only once and it is used by A* pathfinder to determine the shortest path between the set of inspection targets and UAVs.   
\item \textit{Routing solver service} - the service receives inspection targets and UAV locations and uses OR-Tools to determine which UAV should inspect which target. It creates asynchronous requests to the A* service to get the shortest path for each combination of UAV and target. Then, it builds the distance matrix and, based on path length, finds a near-optimal solution to the vehicle routing problem. It returns the path for each UAV as a set of waypoints containing locations in latitude and longitude. 
\item \textit{A* pathfinder service} - the service requests a graph from the graph service and performs the A* algorithm on the targets and UAVs locations received from the routing solver service. A* algorithm is implemented from the NetworkX library. It returns the shortest path between a target and a UAV and returns the total path distance as well as the distance between each path segments with corresponding node locations.   
\end{itemize}

\subsection{Deployment Pipeline and DevOps tools}
In order to deploy the application automatically every time the changes are made, we set up a pipeline using DevOps tools and technologies. Deployment pipeline consists of different stages for application building, deploying, and testing. We also enabled security scanning for code vulnerabilities and monitoring. To store the code, enable collaboration, and configure the pipeline, we used GitLab \cite{b25}.  

Microservices tend to be independent and isolated from each other. The most used and convenient way to isolate services is using container technology. Containers package up the application code and all its dependencies so the application can run quickly and reliably in any computing environment. Docker is the most popular and widely used container technology. Containers run from Docker images configured in Dockerfile. Docker image is a lightweight, standalone, executable package of software that includes everything needed to run an application: code, runtime, system tools, system libraries, and settings. Containers always run the same, regardless of the operating system and infrastructure. Compared with virtual machines, containers isolate application software from its environment but do not include a full copy of an operating system. Therefore, they take up less space and are much faster \cite{b18}. Each LiMiC1.0 service is a Python package containing multiple modules and the Dockerfile. Docker images are built on the Python base image and run the application in a virtual environment after installing all the dependencies.     

\begin{figure}[htbp]
\centerline{\includegraphics[scale=0.5]{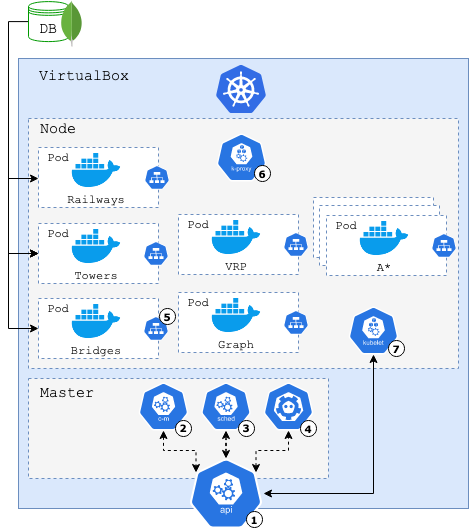}}
\caption{Application deployment in Kubernetes cluster}
\label{fig2}
\end{figure}

In LiMiC1.0, each service runs in a separated container and contains its Dockerfile. The whole process is configured using GitLab. Gitlab has built-in tools for software continuous integration and continuous deployment (CI/CD) which are used to run the pipeline. The codebase is hosted in the GitLab repository and contains a gitlab-ci.yml file describing the pipeline stages. Each service contains its yml file where the deployment and service port is configured. Every time the changes are pushed to the repository, GitLab runners execute scripts defined in the gitlab-ci.yml file and create a cluster deployments based on the service's yml file. We decided to automatize the LiMiC1.0 deployment to the server in the Kubernetes cluster as a staging environment. Figure \ref{fig2} shows the application deployed to the Kubernetes cluster.

Kubernetes is an open-source system for automating deployment, scaling, and management of containerized applications. Kubernetes cluster consists of master and worker machines. Worker machines, also called nodes, run containerized applications in pods while the master manages workers and makes sure that the cluster is working in a configured way. The master node contains an API server which is the cluster entry point. API server \circled{1} runs a process that enables communication with the cluster. To visualize and manage all the cluster applications we can use the dashboard, as a web user interface, communicating with the API server. Otherwise, we can use API to communicate with the API server from the script or simply we can manage the cluster by writing commands in the terminal using a command-line tool. The master node keeps track of a cluster by running a controller manager. Controller manager \circled{2} is responsible for checking if all the nodes and pods are running and, if one goes down, the controller manager replaces it with a new one. Scheduler \circled{3} decides on which node a newly created pod will run, based on the available and required resources. Etcd key-value storage \circled{4} in the master node keeps status data about the nodes. Master nodes and worker nodes communicate through a virtual network. The virtual network assigns an internal IP address to each pod. Pods communicate through the services \circled{5} which contain permanent IP addresses. In case a pod restarts it will keep the service with the same address. Each service also provides a load balancing. Network rules for allowing the communication through the services are maintained by Kube proxy \circled{6} running on each worker node. Kubelet \circled{7} agent runs on each node assuring that containers run in pods as described in pods specifications. Container runtime is responsible for running the containers \cite{b19}. 

We decided to set up a Kubernetes cluster on a server running Linux Ubuntu 18.04.5 LTS using Minikube. Minikube allows you to set up a local cluster with the master process and worker process running on one node. We use it as a staging environment to deploy and test our application since the application development is still in an early stage and does not have any end users. Minikube creates a virtual machine VirtualBox on a local machine and the node runs in the VirtualBox. Kubectl is a command-line tool we use to interact with the cluster. 

\begin{figure}[htbp]
\centerline{\includegraphics[scale=0.45]{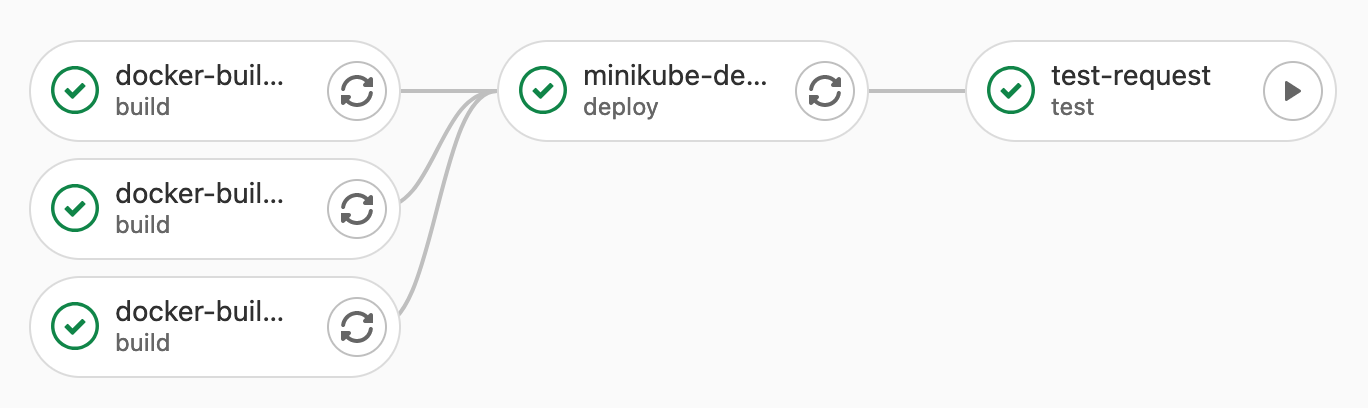}}
\caption{Application deployment pipeline}
\label{fig3}
\end{figure}

To automatize the deployment, we integrated the cluster with GitLab by providing a cluster API and token. To be able to run the CI/CD jobs, we installed the GitLab runner on the cluster. In the yml file, we configured three pipeline stages. In the first stage, the Gitlab runner runs a Docker container for each service based on the Docker image. Inside that container, we build a new image based on the corresponding service's Dockerfile and push it to the Gitlab container registry. The processes run in parallel for each service to speed up the creation of images. In the second stage, we create the deployments and services in the cluster based on the configuration file for each service. In the configuration file, we provide the path to the container registry where the images are stored in the previous stage and set the desired number of pod replicas for each deployment. We configure the ports where the services can be reached. We also enable monitoring through GitLab using Prometheus. The third stage is testing. We test if the services are responsive and if the returned result is correct. The test contains several inspection targets and UAV locations with calculated paths in between. The same targets and UAV locations are sent to the deployed system and the solution is compared with the correct one. The test is delayed for a minute after the deployment stage finishes successfully, to allow the Kubernetes to run all the pods following the yml configuration. The pipeline configuration file also includes a GitLab template for code vulnerability scanning i.e., Static Application Security Testing (SAST). The SAST checks for potentially dangerous attributes in a class, or unsafe code that can lead to unintended code execution and searches for vulnerabilities to cross-site scripting attacks that can be leveraged to unauthorized access to session data. The report can be downloaded after the pipeline executes correctly and the results are sorted by the priority of the vulnerability. In case we are merging changes from another branch to the master branch, GitLab will find the vulnerabilities in the new code that are different from the code in the master branch and report them. GitLab also offers visualization of the pipeline stages where we can see the progress and check the logs for the executed processes. The successful pipeline execution is shown in Figure \ref{fig3}. On every push to the GitLab repository, the code is scanned for the vulnerabilities, the services are containerized, automatically deployed to the Kubernetes cluster, and tested. The controller manager assures that desired number of pods is always running. Pod logs are available in GitLab for both GitLab-managed applications and services.  

\begin{figure}[htbp]
\centerline{\includegraphics[scale=0.35]{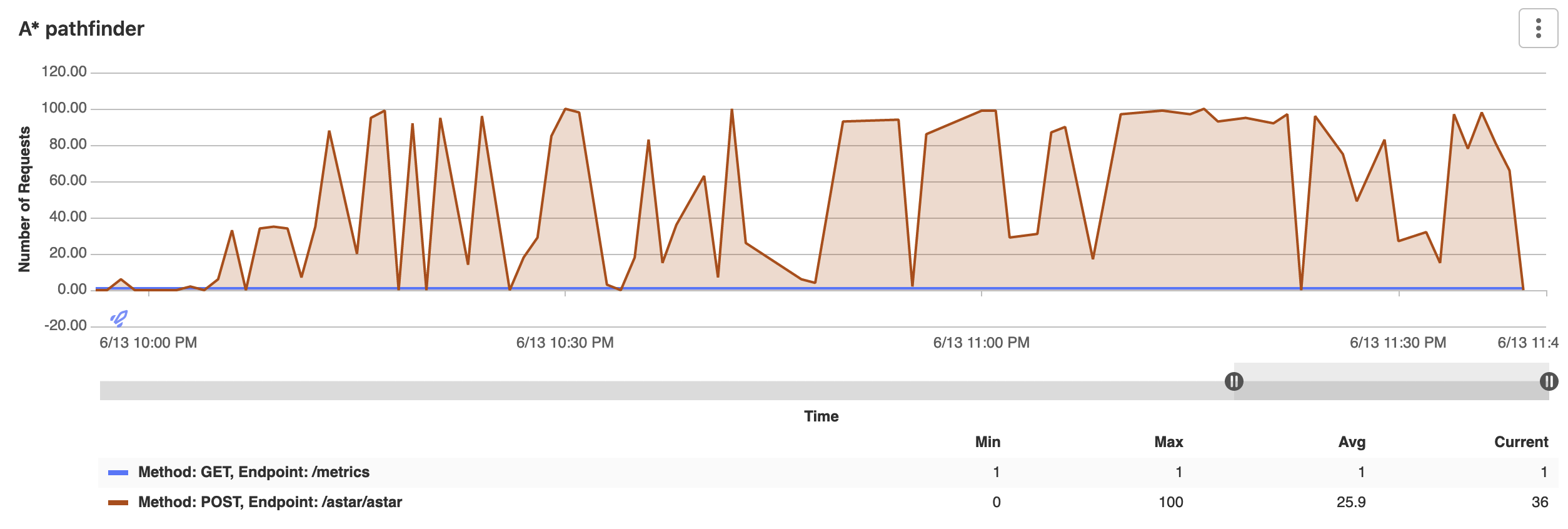}}
\caption{Requests to the A* path finder service}
\label{fig4}
\end{figure}

For monitoring the cluster and the application, we installed an open-source monitoring system Prometheus \cite{b23} through GitLab. For each service, we expose metrics endpoint using Prometheus middleware. In GitLab, we configured a custom dashboard to monitor the number of requests coming to the services. The number of requests coming to the A* pathfinder is visualized in Figure \ref{fig4}. Requests visualized in red are coming from the routing solver while requests in blue are Prometheus requests for scraping metrics. We also monitor the percentage of failed requests. Since GitLab deprecated Prometheus and scheduled it for removal, it was not possible to add alerts. Otherwise, we would be able to add an alert to notify us when there is a certain percentage of failed requests. With Prometheus integration, we also monitor the cluster's CPU and memory usage as well as CPU, memory, and network metrics for each pod in the cluster. That dashboard is set up automatically with the installation of Prometheus.  

\section{Performance Evaluation}
\label{sec:five}
\begin{figure}[H]
\centerline{\includegraphics[scale=0.65]{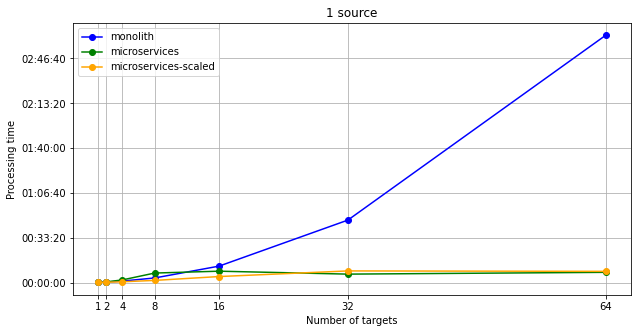}}
\caption{Processing time comparison between monolith and microservice application with 1 source and altering number of targets}
\label{fig5}
\end{figure}

\begin{figure}[H]
\centerline{\includegraphics[scale=0.65]{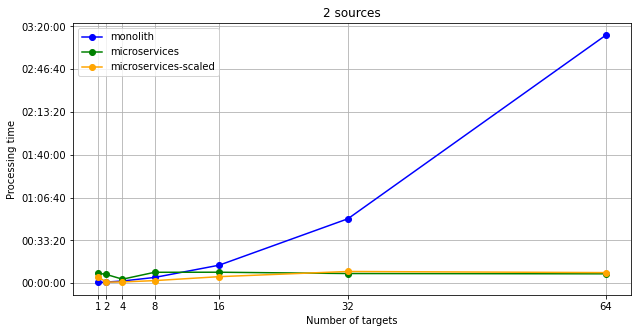}}
\caption{Processing time comparison between monolith and microservice application with 2 sources and altering number of targets}
\label{fig6}
\end{figure}

\begin{figure}[htbp!]
\centerline{\includegraphics[scale=0.65]{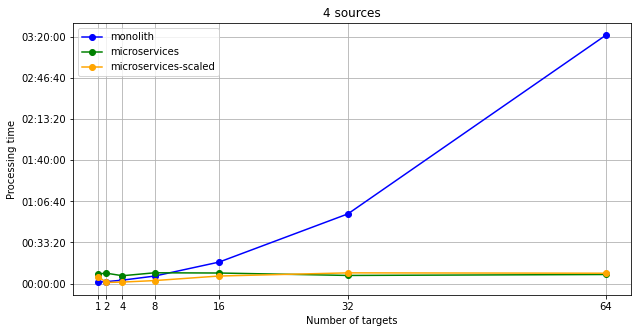}}
\caption{Processing time comparison between monolith and microservice application with 4 sources and altering number of targets}
\label{fig7}
\end{figure}

\begin{figure}[htbp]
\centerline{\includegraphics[scale=0.65]{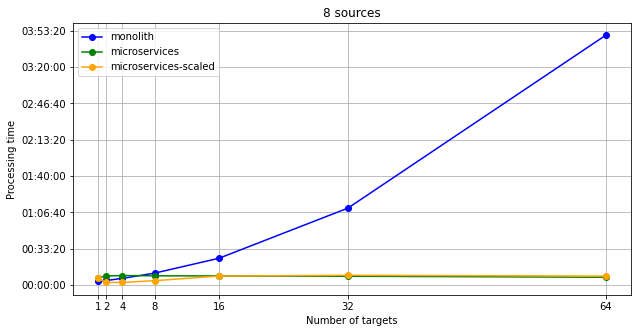}}
\caption{Processing time comparison between monolith and microservice application with 8 sources and altering number of targets}
\label{fig8}
\end{figure}

\begin{figure}[htbp]
\centerline{\includegraphics[scale=0.65]{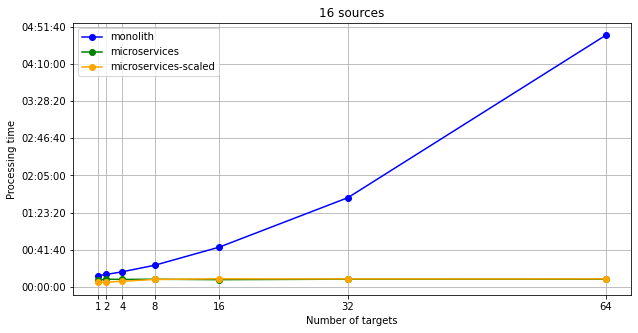}}
\caption{Processing time comparison between monolith and microservice application with 16 sources and altering number of targets}
\label{fig9}
\end{figure}

\begin{figure}[htbp]
\centerline{\includegraphics[scale=0.65]{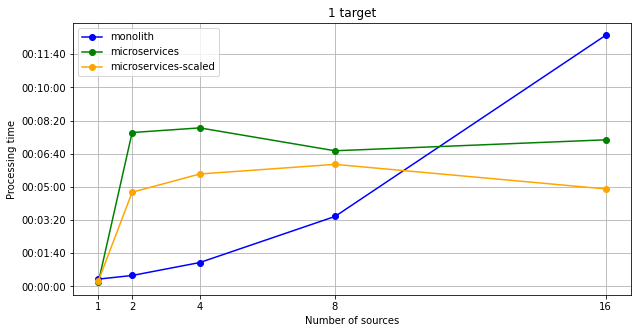}}
\caption{Processing time comparison between monolith and microservice application with 1 target and altering number of sources}
\label{fig10}
\end{figure}

\begin{figure}[htbp]
\centerline{\includegraphics[scale=0.65]{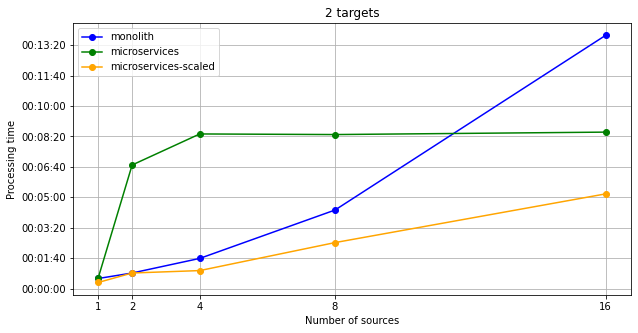}}
\caption{Processing time comparison between monolith and microservice application with 2 targets and altering number of sources}
\label{fig11}
\end{figure}

\begin{figure}[htbp]
\centerline{\includegraphics[scale=0.65]{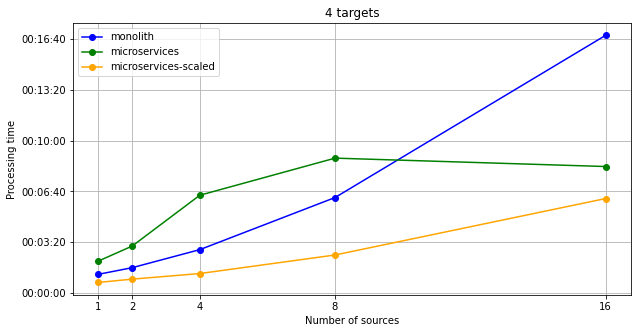}}
\caption{Processing time comparison between monolith and microservice application with 4 targets and altering number of sources}
\label{fig12}
\end{figure}

\begin{figure}[htbp]
\centerline{\includegraphics[scale=0.65]{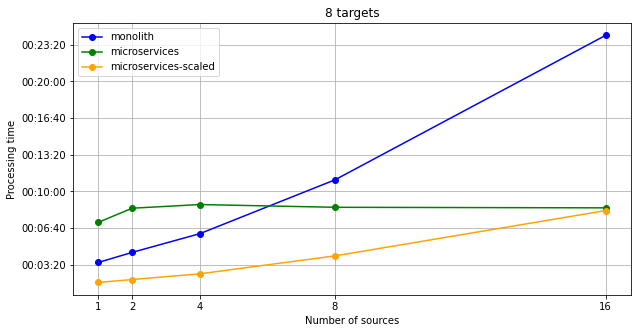}}
\caption{Processing time comparison between monolith and microservice application with 8 targets and altering number of sources}
\label{fig13}
\end{figure}

\begin{figure}[htbp]
\centerline{\includegraphics[scale=0.65]{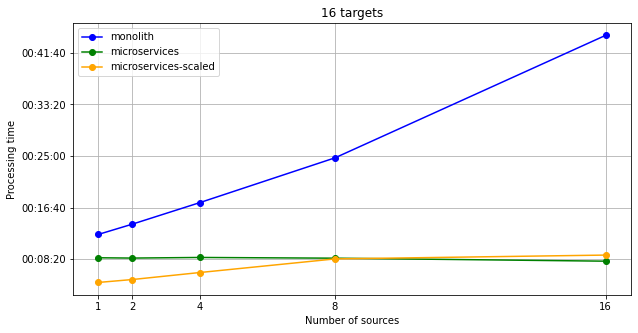}}
\caption{Processing time comparison between monolith and microservice application with 16 targets and altering number of sources}
\label{fig14}
\end{figure}

\begin{figure}[htbp]
\centerline{\includegraphics[scale=0.7]{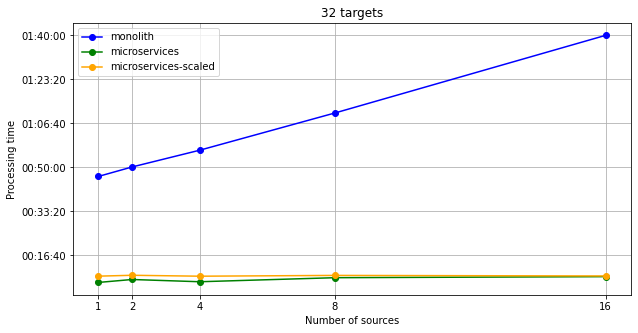}}
\caption{Processing time comparison between monolith and microservice application with 32 targets and altering number of sources}
\label{fig15}
\end{figure}

\begin{figure}[htbp]
\centerline{\includegraphics[scale=0.65]{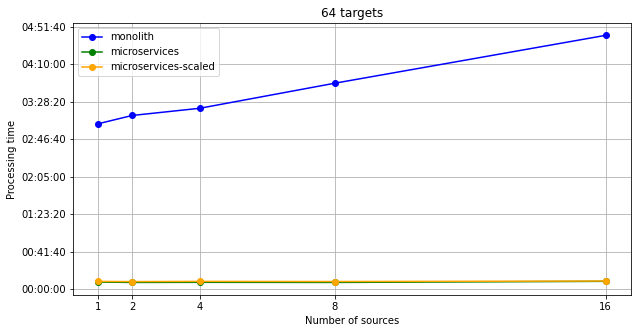}}
\caption{Processing time comparison between monolith and microservice application with 64 targets and altering number of sources}
\label{fig16}
\end{figure}

To evaluate our application performance, we deployed it automatically through the GitLab CI/CD to the Minikube Kubernetes cluster. We compared the performance with the monolithic version of the application. To assure the same running environment, we created a Docker image for our monolith application and deployed it to the same cluster within a new namespace. We tested the performance by sending requests with different sets of sources and targets and measuring the time to process the requests. Sources represent UAV locations and targets are power tower locations. We chose the locations contained in the graph randomly. The data contains power tower locations in Denmark and we suppose that UAVs are located on the power tower locations on their start. We exponentially increment the number of sources from 1 to 16, and the number of targets from 1 to 64. One hundred requests were generated for each combination of sources and targets, thus 3500 requests in total. The same requests were made to both systems. Requests to LiMiC1.0 were made for a different number of pod deployments. First, we deployed only one pod per service and measured the processing time. Then we scaled the deployment with ten A* pathfinder pod replicas to take the advantage of asynchronous communication between vehicle routing solver and A* pathfinder services. The processing time comparison between monolithic LiMiC and microservices LiMiC1.0 is shown in graphs in Figures \ref{fig5} to \ref{fig16}. Processing time for microservices application without scaling is shown in green, with ten A* pathfinder pod replicas is shown in yellow, and the monolithic application processing time is shown in blue. Figures \ref{fig5} to \ref{fig9} show processing time for a different number of sources, i.e. UAVs, while altering the number of targets. Figures \ref{fig10} to \ref{fig16} show processing time for a different number of targets while altering the number of sources. Both options were visualized to provide a better analysis of the system's performance. 

In all figures, we see that with the monolithic application, the processing time extends proportionally with the number of sources and targets while with microservices the processing time stabilizes and stops raising. For the microservices application, the overhead of internal communication between the services adds in processing time for two to eight sources visiting one target, which is best visible in Figure \ref{fig10}, where both scaled and not scaled microservice application performs slower than the monolith. The same is visible in Figures \ref{fig6} to \ref{fig8}, where the path is calculated for sources to reach only one target. For a high number of sources, as in Figure \ref{fig9}, microservices perform faster for each number of targets. As the number of targets is raising, monolithic application processing time slows down, and microservices start outperforming the monolith, especially when scaled, as seen in Figures \ref{fig11}, \ref{fig12}, and \ref{fig13}. For 16 targets, LiMiC1.0 performs significantly better as shown in \ref{fig14}. By increasing the number of targets even more the processing time becomes near constant as seen in Figures \ref{fig15} and \ref{fig16}. If we take the highest number of sources and targets and calculate the average time it takes to find near-optimal paths for 16 UAVs to visit 64 towers, we calculate that LiMiC1.0, when scaled, can process the request and serve the solution in about 5,32 seconds. Compared with LiMiC which takes 1 minute and 13 seconds on average. When scaled, LiMiC1.0 is slower than the monolithic LiMiC only when two, four, or eight UAVs are chosen for inspection of one target. However, it is unlikely for a user to pick either of the combinations mentioned, since most of the time it is intuitive when visualized, which UAV is the most suitable for the mission inspecting only one target. We also have to take into account that sources and targets are chosen randomly over the country of Denmark. Therefore they do not always represent a realistic set and sometimes it was not even possible for the solver to find a solution. We expect that the average processing time for a realistic set of sources and targets is slightly better. However, this experiment provided an insight into applications' performance and showed a positive impact of migrating the route calculation algorithm to microservices. The performance of the microservices application could be further improved by deploying to a more powerful cluster and taking the advantage of automatic scaling.

\section{Conclusion \& Future Work}
\label{sec:six}
To increase safety and reduce costs while inspecting the infrastructure, we developed an application for routing and scheduling UAVs to fly near the infrastructure. In this paper, we redesigned the application architecture to speed up future development, enable collaboration between developers, automate the deployment, and most important to speed up the route calculation. We implemented the application in microservice architecture and configured deployment pipeline stages using GitLab. We automatically trigger the Docker image creation, push the images to the container registry, scan the code for security vulnerabilities, deploy the application to the Minikube Kubernetes cluster and test the services' responsiveness. Also, we check if the results calculated are correct. We test the performance by comparing monolithic application and microservice application processing times. Microservices process requests faster, having the advantage of pod scaling and asynchronous communication between the services. Monolith application processing time exponentially rises when calculating paths for an increasing number of UAVs and power towers to inspect. By refactoring the calculation algorithm, the processing time is shorter and stabilizes after reaching a certain number of sources and targets.       

To have a fully operational application for infrastructure inspection, we aim at implementing the additional services. In some areas, UAV flights are not permitted and we want to avoid these zones. While creating the graph, we will detect the infrastructure locations laying in the forbidden areas and reroute around them to assure flight safety. The application has to be able to communicate with UAV and send the navigation waypoints calculated in the routing solver service. Therefore, we aim at implementing a message broker service to manage the communication between UAVs and the application. UAVs should be able to send data, including inspection images, through the same message broker. One of the additional services would estimate the UAVs' location to visualize their location on the web interface in every moment while the mission is in progress, even if the connectivity is not constant. We will implement storage for planned and executed missions to be able to retrieve the mission data after its execution. The application would also contain image analysis services to detect faulty infrastructure using machine learning models as well as image storage. Another important feature is a service for 3D infrastructure reconstruction which would be used for path planning around the power towers and bridges. An idea worth considering for future work would be to implement the weather prediction service and use it to determine if the weather in the area allows a safe flight. The microservice architecture we implemented and presented in this paper will allow us to add mentioned features easily as independent services. Concerning the deployment strategy, we aim at deploying the application to production to a multi-node Kubernetes cluster in the public cloud and test the application's load.           

\section{Acknowledgements}
This project has received funding from European Union's Horizon 2020 Research and Innovation Programme under Grant Agreement No 861111, Drones4Safety.

\vspace{12pt}

\end{document}